\documentclass[conference]{IEEEtran}
\IEEEoverridecommandlockouts
\usepackage{amsmath,amssymb,amsfonts}
\usepackage{amsthm}
\usepackage[ruled,vlined,linesnumbered]{algorithm2e}
\usepackage{graphicx}
\usepackage{textcomp}
\usepackage{xcolor}
\usepackage{subfigure}
\usepackage{booktabs}
\usepackage{multirow}
\usepackage{makecell}
\usepackage{hyperref}
\def\BibTeX{{\rm B\kern-.05em{\sc i\kern-.025em b}\kern-.08em
    T\kern-.1667em\lower.7ex\hbox{E}\kern-.125emX}}
    
\SetKw{And}{and}
\SetKw{Or}{or}
\SetKw{True}{true}
\SetKw{False}{false}
\SetKw{Break}{break}
\SetKw{To}{to}
\theoremstyle{definition}

\theoremstyle{plain}
\newtheorem{Lemma}{\textbf{Lemma}}

\theoremstyle{definition}
\newtheorem{Definition}{\textbf{Definition}}

\begin{document}

\title{Improving Reverse~$k$~Nearest Neighbors Queries
}


\author{\IEEEauthorblockN{Lixin Ye
}
\IEEEauthorblockA{\textit{School of Computer Science, China University of Geosciences}}
}

\maketitle

\begin{abstract}
The reverse $k$ nearest neighbor query finds all points that have the query point as one of their $k$ nearest neighbors, where the $k$NN query finds the $k$ closest points to its query point.
Based on conics, we propose an efficent R$k$NN verification method. By using the proposed verification method, we implement an efficient R$k$NN algorithm on VoR-tree, which has a computational complexity of $O(k^{1.5}\cdot log\,k)$.
The comparative experiments are conducted between our algorithm and other two state-of-the-art R$k$NN algorithms.
The experimental results indicate that the efficiency of our algorithm is significantly higher than its competitors.
\end{abstract}

\begin{IEEEkeywords}
R$k$NN, conic section, Voronoi, Delaunay
\end{IEEEkeywords}

\section{Introduction}
As a variant of nearest neighbor (NN) query, RNN query is first introduced by Korn and Muthukrishnan \cite{DBLP:conf/sigmod/KornM00}.
A direct generalization of NN query is the reverse $k$ nearest neighbors (R$k$NN) query, where all points having the query point as one of their $k$ closest points are required to be found.
Since its appearance, R$k$NN has received extensive attention \cite{six,TPL,FINCH,InfZone,SLICE,VoR} and been prominent in various scientific fields including machine learning, decision support, intelligent computation and geographic information systems, etc.

At first glance, R$k$NN and $k$NN queries appear to be equivalent, meaning that the results for R$k$NN and $k$NN may be the same for the same query point.
However, R$k$NN is not as simple as it seems to be.
It is a very different kind of query from $k$NN, although their results are similar in many cases.
So far, R$k$NN is still an expensive query for its computational complexity at $O(k^2)$ \cite{SLICE}, whereas the computational complexity of $k$NN queries has been reduced to $O(k\cdot log\, k)$ \cite{VoR}.

In order to solve the RNN/R$k$NN problem, a large number of approaches have been proposed.
Some early methods \cite{MVZ,DBLP:conf/sigmod/KornM00,rdnn-tree} speed up RNN/R$k$NN queries by pre-computation.
Their disadvantage is that it is difficult to support queries on dynamic data sets.
Therefore, many R$k$NN algorithms without pre-computation are proposed.

Most existing non-pre-computation R$k$NN algorithms have two phases: the filtering phase and the refining phase (also known as the pruning phase and the verification phase).
In the pruning phase, the majority of points that do not belong to R$k$NN should be filtered out.
The main goal of this phase is to generate a candidate set as small as possible.
In the verification phase, each candidate point should be verified whether it belongs to the R$k$NN set or not.
For most algorithms, the candidate points are verified by issuing $k$NN queries or range queries, which are very computational expensive.
The state-of-the-art R$k$NN technique SLICE, provides a more efficient verification method with a computational complexity of $O(k)$ for one candidate. The size of the candidate set of SLICE varies form $2k$ to $3.1k$. However, it is still time consuming to perform such a verification for each candidate point.

There seems to be a consensus in the past studies that for an R$k$NN technique, the number of verification points cannot be smaller than the size of the result set. Such an idea, however, limits our understanding of the R$k$NN problem. 
Hence we amend our thought and come up with a conjecture that  whether a point could be directly determined as belonging to the R$k$NN set according to its location.
Given the query point $q$, our intuition tells us that if a point $p$ is closer to $q$ than a point $p^+$ belonging to the R$k$NNs of $q$, then $p$ is highly likely to also belong to the R$k$NN of $q$.
Conversely, if $p$ is further away from $q$ than a point $p^-$ that does not belong to the R$k$NN set of $q$, then $p$ is probably not a member of the R$k$NN set.
Along with this idea, we further study and obtain a set of verification methods for R$k$NN queries.
Based VoR-tree, we use this veirification method implement an efficient R$k$NN algorithm, which out performs most mainstream algorithms.

\begin{table}[htbp]
\centering
\caption{Comparison of computational complexity}
\label{table:comlexity}
\begin{tabular}{@{}lccc@{}}
\toprule
\textbf{Operation} & \textbf{VR-R$k$NN}& \textbf{SLICE}& \textbf{Our approach}\\ \midrule
 Generate candidates & $O(k\cdot log\, k)$ & $O(k\cdot log\, k)$ & $O(k\cdot log\, k)$\\
 Verify a candidate & $O(k\cdot log\, k)$ & $O(k)$ & $O(k\cdot log\, k)$\\
$\vert$Verified candidates$\vert$ & \makecell[c]{$O(k)$\\(=6$k$)} & \makecell[c]{$O(k)$\\ (2$k$$\sim$3.1$k$)} &\makecell[c]{ $O(\sqrt{k})$\\($\leq 7.1\sqrt{k}$)}\\
Overall & $O(k^2\cdot log\, k)$ & $O(k^2)$ & $O(k^{1.5}\cdot log\, k)$ \\ \bottomrule
\end{tabular}
\end{table}

Table~\ref{table:comlexity} shows the comparison of computational complexity among VR-R$k$NN , SLICE and our approach.
It can be seen that the bottleneck of both VR-R$k$NN and SLICE is the verification phase. The computational complexity  of verifying a candidate of our approach is $O(k\cdot log\, k)$, which is higher than that of SLICE.
However, the number of candidates verified by our approach  is only about $7.1\sqrt{k}$, which is much less than that of SLICE. In addition, the overall computational complexity of our approach is much lower than that of SLICE.

The rest of the paper is organized as follows. In Section~2, we introduce the major related work of R$k$NN since its appearance. In Section~3, we formally define the R$k$NN problem and introduce the concepts and knowledge related to our approach. Our approach and its principles are described in section~4. Section~5 provides a detailed theoretical analysis. Experimental evaluation is demonstrated in Section~6. The last two sections are conclusions and acknowledgements.

\section{Related work}
\subsection{RNN-tree}
Reverse nearest neighbor (RNN) queries are first introduced by Korn and Muthukrishnan where RNN queries are implemented by preprocessing the data \cite{DBLP:conf/sigmod/KornM00}.
For each point $p$ in the database, a circle with $p$ as the center and the distance from $p$ to its nearest neighbor as the radius is pre-calculated and these circles are indexed by an R-tree.
The RNN set of a query point $q$ includes all the points whose circle contains $q$. 
With the R-tree, the RNN set of any query point can be found efficiently.
Soon after, several techniques \cite{DBLP:conf/icde/YangL01,DBLP:conf/ideas/LinNY03} are proposed to improve their work.
\subsection{Six-regions}
Six-regions \cite{six} algorithm, proposed by Stanoi et al., is the first approach that does not need any pre-computation.
They divide the space into six equal segments using six rays starting at the query point, so that the angle between the two boundary rays of each segment is 60$^\circ$.
They suggest that only the nearest neighbor (NN) of the query point in each of the six segments may belong to the RNN set.
It firstly performs six NN queries to find the closest point of the query point $q$ in each segments. Then it launches an NN query for each of the six points to verify $q$ as their NN. Finally the RNN of $q$ is obtained.

Generalizing this theory to R$k$NN queries leads to a corollary that,
only the members of $k$NN of the query point in each  segment have the possibility of belonging to the R$k$NN set.
This corollary is widely adopted in the pruning phase of several R$k$NN techniques.


\subsection{TPL}
TPL \cite{TPL}, proposed by Tao et al., is one of the prestigious algorithms for RkNN queries.
This technique prunes the space using the bisectors between the query point and other points. 
The perpendicular bisector is denoted by $B_{p:q}$. $B_{p:q}$ is between a point $p$ and the query point $q$.
$B_{p:q}$ divides the space into two half-spaces. The half-space that contains $p$ is denoted as $H_{p:q}$. Another one is denoted as $H_{q:p}$.
If a point $p'$ lies in $H_{p:q}$, $p'$ must be closer to $p$ than to $q$. Then $p'$ cannot be the RNN of $q$ and we can say that $p$ prunes $p'$. If a point is pruned by at least $k$ other points, then it cannot belong to the R$k$NN of $q$.
An area that is the intersection of any combination of $k$ half-spaces can be pruned.
The total pruned area corresponds to the union of pruned regions by all such possible combinations of $k$ bisectors (total $\dbinom{m}{k}$ combinations).
TPL also uses an alternative computational cheaper pruning method which has a less pruning power. All the points are sorted by their Hilbert values.
Only the combinations of $k$ consecutive points are used to prune the space  (total $m$ combinations).

\subsection{FINCH}
FINCH is another famous R$k$NN algorithm proposed by Wu et al. \cite{FINCH}.
The authors of FINCH think that it is too computational costly to use $m$ combinations of $k$ bisectors to prune the points.
They utilize a convex polygon that approximates the unpruned region to prune the points instead of using bisectors. All points lying outside the polygon should be pruned.
Since the containment can be achieved in logarithmic time for convex polygons, the pruning of FINCH has a higher efficiency than TPL.
However, the computational complexity of computing the approximately unpruned convex polygon is $O(m^3)$, where $m$ is the number of points considered for pruning. 

\subsection{InfZone}
Previous techniques
can reduce the candidate set to an extent by different pruning methods.
However, their verification methods for candidates are very inefficient. It is quite computational costly to issue an inefficient verification for each point in a candidate set with a size of O($k$).
In order to overcome this issue, a novel R$k$NN technique which is named as InfZone is proposed by Cheema et al. \cite{InfZone}.
The authors of InfZone introduce the concept of influence zone (denoted as $Z_k$), which also can be called R$k$NN region.
The influence zone of a query point $q$ is a region that, a point $p$ belongs to the R$k$NN set of $q$, if and only if it lies in the $Z_k$ of $q$.
The influence zone is always a star-shaped polygon and the query point is its kernel point.
A number of properties are detailed. These properties are aimed to shrink the number of points which are crucial to compute the influence zone.
They propose an influence zone computing algorithm with a computational complexity of $O(k\cdot m^2)$, where $m$ is the number of points accessed during the construction of the influence zone.
Every points that lies inside the influence zone are accessed in the pruning phase, since they cannot be ignored during the construction of the influence zone.
Namely, all the potential members of the R$k$NN are accessed during the pruning phase.
Hence, for monochromatic R$k$NN queries, InfZone does not require to verify the candidates.
It is indicated that the expected size of R$k$NN set is $k$. 
Evidently, the size of R$k$NN must not be greater than $m$, i.e., $k \leq m$.
Therefore, the computational complexity of InfZone must be no less than $O(k^3)$.

\subsection{SLICE}
SLICE \cite{SLICE} is the state-of-the-art approach for R$k$NN queries.
In recent years,  several well-known techniques \cite{six} have been proposed to address the limitations of half-space pruning\cite{TPL} (e.g., FINCH \cite{FINCH}, InfZone \cite{InfZone}).
While few researcher carries out further research based on the idea of Six-regions.
Yang et al. suggests that the regions-based pruning approach of Six-regions has great potential and proposed an efficient R$k$NN algorithm SLICE \cite{SLICE}.
SLICE uses a more powerful and flexible pruning approach that prunes a much larger area as compared to Six-regions with almost similar computational complexity. 
Furthermore, it significantly improves the verification phase by computing a list of significant points for each segment.
These lists are named as $sigList$s.
Each candidate can be verified by accessing $sigList$ instead of issuing a range query.
Therefore, SLICE is significantly more efficient
than the other existing algorithms.

\subsection{VR-R$k$NN}
For most R$k$NN algorithms, data points are indexed by R-tree \cite{R-TREE}.
However, R-tree is originally designed primarily for range queries.
Although some approaches \cite{best-first,TPL,ANNQ,skyline} are proposed afterwards to make it also suitable for NN queries and their variants: the NN derived queries are still disadvantageous. When answering an NN derived query, all nodes in the R-tree intersecting with the local neighborhood (Search Region) of the query point need to be  accessed to find all the members of the result set.
Once the candidate set of the query is large, the cost of accessing the nodes can also become very large.
In order to improve the performance of R-tree on NN derived queries, Sharifzadeh and Shahabi proposes a composite index structure composed of an R-tree and a Voronoi diagram, and named it as VoR-Tree \cite{VoR}.
VoR-Tree benefits from both the neighborhood exploration capability of Voronoi diagrams and the hierarchical structure of R-tree.
By utilizing VoR-tree, they propose VR-R$k$NN to answer the R$k$NN query.
Similar to the filter phase of Six-regions \cite{six}, Vor-R$k$NN divides the space into 6 equal segments and selects $k$ candidate points from each segment to form a candidate set of size 6$k$. During the refining phase, each candidate point is verified to be a member of the R$k$NN through issuing a $k$NN query (VR-$k$NN).
The expected computational complexity of VR-R$k$NN is $O(k^2\cdot log\, k)$

\section{Preliminaries}
\subsection{Problem definition}
\begin{Definition}
\textbf{Euclidean Distance:} Given two points $A=\{a_1, a_2,$ $...,a_d\}$ and $B=\{b_1, b_2,...,b_d\}$ in $\mathbb{R}^d$, the Euclidean distance between $A$ and $B$, $dist(A, B)$, is defined as follows:
\begin{align}
dist(A,B)=\sqrt{\sum_{i=1}^d(a_i-b_i)^2}.
    \label{eq:dist}
\end{align}
\end{Definition}

\begin{Definition}
\textbf{$k$NN Queries:}
A $k$NN query is to find the $k$ closest points to the query point from a certain point set.
Mathematically, this query in Euclidean space can be stated as follows. 
Given a set $P$ of points in $\mathbb{R}^d$ and a query point $q \in \mathbb{R}^d$,
\begin{align}
k{\rm NN}(q) = \{p \in P\;|\;dist(p, q) \leq dist(p_k, q)\} \nonumber\\ {\rm where}\;p_k\;{\rm is\;the}\;k {\rm th\;closest\;point\;to}\;q\;{\rm in}\;P.\;
    \label{eq:knn}
\end{align}
\end{Definition}

\begin{Definition}
\textbf{R$k$NN Queries:} A R$k$NN query retrieves all the points that have the query point as one of their $k$ nearest neighbors from a certain point set.
Formally, given a set $P$ of points in $\mathbb{R}^d$ and a query point $q \in P$, the R$k$NN of $q$ in $P$ can be defined as
\begin{align}
{\rm R}k{\rm NN}(q) = \{p \in P\;|\;q \in k{\rm NN}(p) \}.
    \label{eq:rknn}
\end{align}
\end{Definition}

\subsection{Voronoi diagram \& Delaunay graph}
\begin{figure}[htbp]
    \centering
    \includegraphics[width=0.48\textwidth]{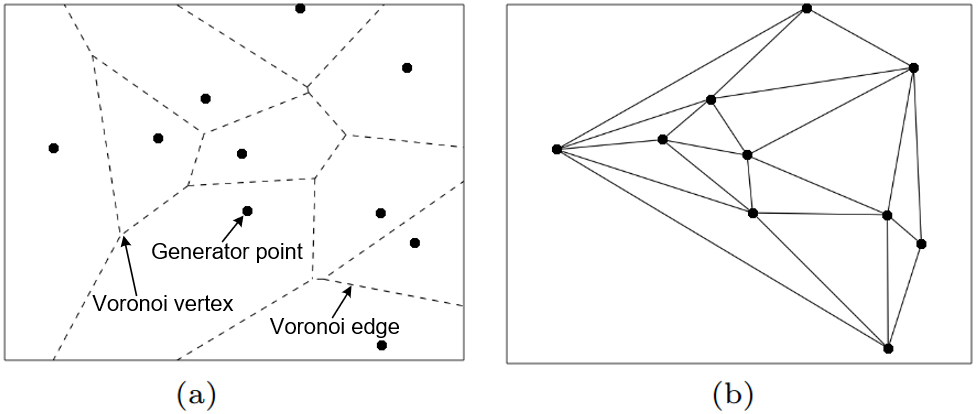}
    \caption{a) Voronoi Diagram, b) Delaunay Graph}
    \label{fig:vd_dg}
\end{figure}
Voronoi diagram \cite{DBLP:reference/gis/ShahabiS17}, proposed by Rene Descartes in 1644, is a spatial partition structure widely applied in many science domains, especially spatial database and computational geometry. In a Voronoi diagram of $n$ points, the space is divided into $n$ regions corresponding to these points, which are called Voronoi cells.
For each of these $n$ points, the corresponding Voronoi cell consists of all locations closer to that point than to any other.
In other words, each point is the nearest neighbor of all the locations in its corresponding Voronoi cell.
Formally, the above description can be stated as follows.
\begin{Definition}
\textbf{Voronoi cell \& Voronoi diagram:}
Given a set $P$ of $n$ points, the Voronoi cell of a point $p \in P$, denoted as $V(P, p)$ or $V(p)$ for short, is defined as Equation~\eqref{voronoi_cell}
\begin{align}
V(P, p)=\{q\;|\;
\forall p' \in P\setminus\{p\}:\;dist(p,q) \leq dist(p',q)\}
    \label{voronoi_cell}
\end{align}
and the Voronoi diagram of $P$, denoted as $VD(P)$, is defined as Equation~\eqref{voronoi_diagram}.
\begin{align}
VD(P)=\{V(P, p)\;|\; p \in P\}
    \label{voronoi_diagram}
\end{align}
The Voronoi diagram of a certain set $P$ of points, $VD(P$), is unique.
\end{Definition}

\begin{Definition}
\textbf{Voronoi neighbor:}
Given the Voronoi diagram of $P$,
for a point $p$, its Voronoi neighbors are the points in $P$ whose Voronoi cells share an edge with $V(P,q)$. It is denoted as $VN(P,q)$ or $VN(q)$ for short.
Note that the nearest point in $P$ to $p$ is among $VN(q)$.
\end{Definition}

\begin{Lemma}
\label{lemma:nn}
 Let $p_k$ be the $k$-th nearest neighbor of $q$,  then $p_k$ is a Voronoi neighbor of
 at least one point of the $k-1$ nearest neighbors of $q$ (where $k>1$).
\end{Lemma}
\begin{proof}
See \cite{VoR}.
\end{proof}

\begin{Lemma}
For a Voronoi diagram, the expected number of Voronoi neighbors of a generator point does not exceed 6. 
\label{lemma:not_exceed_6}
\end{Lemma}
\begin{proof}
Let $n$, $n_e$ and $n_v$ be the number of generator points, Voronoi edges and Voronoi vertices of a Voronoi diagram in $\mathbb{R}^2$, respectively, and assume $n \geq 3$. According to Euler's formula,
\begin{align}
n+n_v-n_e=1
    \label{eq6}
\end{align}
Every Voronoi vertex has at least 3 Voronoi edges and each Voronoi edge  belongs to two Voronoi vertices. Hence the number of Voronoi edges is not less than $3(n_v+1)/2$, i.e.,
\begin{align}
n_e \geq \frac{3}{2}(n_v+1)
    \label{eq7}
\end{align}
According to Equation~\eqref{eq6} and Equation~\eqref{eq7}, the following relationships holds:
\begin{align}
n_e \leq 3n-6
    \label{eq8}
\end{align}
When the number of generator points is large enough, the average number of Voronoi edges  per  Voronoi  cell  of a Voronoi diagram in $\mathbb{R}^d$  is a constant value depending only on $d$. When $d$ = 2, every Voronoi edge is shared by two Voronoi Cells. Hence the average number of Voronoi edges per Voronoi cell does not exceed 6, i.e., $2 \cdot n_e/n \leq 2(3n-6)/n= 6-12/n \leq 6$.
\end{proof}

For set of points $P$, a dual graph of its Voronoi Diagram is the Delaunay graph (denoted as $DG(P)$) \cite{Delaunay1934Sur} of it.
For $P$, its nearest neighbor graph is a subgraph of its Delaunay graph.

\begin{Definition}
\textbf{Delaunay graph distance:}
Given the Delaunay graph $DG(P)$, the Delaunay graph distance between two vertices $p$ and $p'$ of $DG(P)$ is the minimum number of edges connecting $p$ and $p'$ in $DG(P)$. It is denoted as $dist_{DG}(p,p')$.
\end{Definition}
\begin{Lemma}
\label{lemma:dgd}
Given the query point $q$,
if a point $p$ belongs to R$k$NN($q$), then we have
$dist_{DG}(p, q) \leq k$ in Delaunay graph $DG(p)$.
\end{Lemma}
\begin{proof}
See \cite{VoR}.
\end{proof}

\subsection{Conic section}

\begin{Definition}
\textbf{Ellipse:} An ellipse is a closed curve on a plane, such that the sum of the distances from any point on the curve to two fixed points $p_1$ and $p_2$ is a constant $C$.
Formally, it is denoted as $E^c_{p_1: p_2}$ defined as follows:
\begin{align}
    E^c_{p_1: p_2}=\{p\;|\;dist(p,p_1)+dist(p,p_2)=C\}
\end{align}
\end{Definition}

\begin{Definition}
\textbf{Hyperbola:} A hyperbola is a geometric figure such that the difference between the distances from any point on the figure to two fixed points $p_1$ and $p_2$ is a constant $C$.
Formally, it is denoted as $H^c_{p_1: p_2}$defined as follows:
\begin{align}
    H^c_{p_1: p_2}=\{p\;|\;|dist(p,p_1)-dist(p,p_2)|=C\}
\end{align}
\end{Definition}

\section{Methodologies}


\subsection{Verification approach}

\begin{Definition}
\begin{figure}[htbp]
    \centering
    \includegraphics[width=0.3\textwidth]{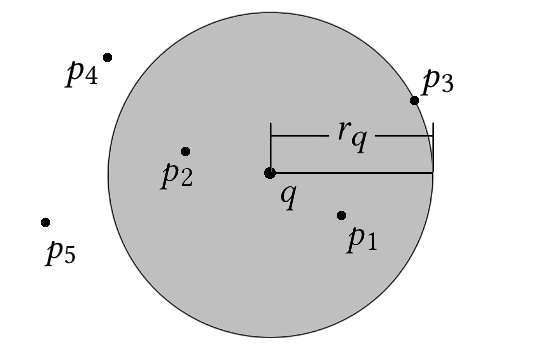}
    \caption{$k$NN region}
    \label{fig:knn_region}
\end{figure}
\textbf{$k$NN region:} Given a query point $q$, the $k$NN region of $q$ is the inner region of $C_{q:dist(q,p_k)}$, i.e., the circle with $q$ as center and $dist(q,p_k)$ as the length of radius, where $p_k$ represents the $k$th closest point to $q$. This region is denoted as $RG${\tiny$_{k{\rm NN}}$}$(q)$. The radius of $RG${\tiny$_{k{\rm NN}}$}$(q)$ is called the $k$NN radius of $q$ and is denoted as $r_q$.

Note that a point $p$ must be one $k$NN($q$) if it lies in $RG${\tiny$_{k{\rm NN}}$}$(q)$, i.e., the $k$NN region of $q$. Conversely, if a point $p'$ lies out of $RG${\tiny$_{k{\rm NN}}$}$(q)$, it cannot be any one of $k$NN($q$).
In Figure~\ref{fig:knn_region}, $q$ is the query point and the gray region  within the circle centered on $q$ represents $RG${\tiny$_{k{\rm NN}}$}$(q)$. As we can see, $p_1$, $p_2$ and $p_3$ lie inside $RG${\tiny$_{k{\rm NN}}$}$(q)$, then we can determine that they belong to $k$NN($q$). while $p_4$ and $p_5$ lie outside. So they are not the members of $k$NN($q$).
\end{Definition}

\begin{Lemma}
\label{lemma_1}
Given a query point $q$,
a point $p$ must be one of R$k$NN($q$) if it satisfies 
\begin{align}
\label{eq_lemma1_0}
dist(p,q) \leq r_p.
\end{align}
Conversely, a point $p'$ cannot be any one of R$k$NN($q$) if it satisfies
\begin{align}
\label{eq_lemma1_1}
dist(p',q) > r_{p'}.
\end{align}
Simply, for a point $p$, if the query point $q$ lies in its $k$NN region, $p$ must be one of R$k$NN($q$), otherwise it must not belong to R$k$NN($q$).
\end{Lemma}
\begin{proof}
The lemma is easily proved by the definition of $k$NN  and R$k$NN, see Equation~\eqref{eq:knn} and Equation~\eqref{eq:rknn}.
\end{proof}
According to Lemma~\ref{lemma_1}, we can determine whether a point $p$ belongs to the R$k$NN of the query point $q$ by calculating the $k$NN region of $p$.
Obviously, $q$ lying in $RG${\tiny$_{k{\rm NN}}$}$(p)$ is a necessary and sufficient condition for $p$ to be one of R$k$NN($q$).
In the refining phase of some R$k$NN algorithms, the candidates are verified by this condition.
In this verification method, $k$NN region is required, so a $k$NN query must be conducted. 
The computational complexity of the state-of-the-art $k$NN algorithm is $O(k\cdot log\,k)$.
Thus, the computational complexity of the verification method based on  Lemma~\ref{lemma_1} is $O(k\cdot log\,k)$.

For most R$k$NN algorithms, the size of candidate set is often several times much as that of the result set.
Therefore, issuing a R$k$NN verification of which the computational complexity is $O(k\cdot log\,k)$ for each candidate is obviously expensive.
In order to reduce the computational cost of the refining phase of R$k$NN queries, we introduce several more efficient verification approaches in the following.

\begin{Lemma}
\label{lemma_2}
Given a query point $q$ and a point $p^+ \in$ R$k$NN($q$), a point $p$ must be one of R$k$NN($q$) if it satisfies
\begin{align}
\label{eq:lemma2_0}
dist(p,q)+dist(p,p^+) \leq r_{p^+}.
\end{align}
\end{Lemma}

\begin{figure}[htbp]
    \centering
    \includegraphics[width=0.25\textwidth]{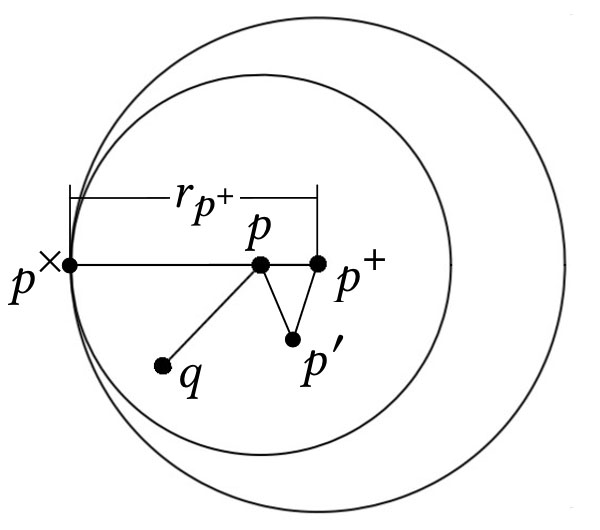}
    \caption{Lemma \ref{lemma_2}}
    \label{fig:lemma2}
\end{figure}

\begin{proof}
As shown in Figure~\ref{fig:lemma2}, the larger circle takes $p^+$ as the center and $r_{p^+}$ as the radius, which represents the $k$NN region of $p^+$.
$L_{p^\times,p^+}$ is a line segment passing through the point $p$ with a length of $r_{p^+}$.
The smaller circle takes $p$ as the center and $dist(p,p^\times)$ as the radius.
Let $p'$ be an arbitrary point inside  $C_{p:dist(p,p^\times)}$, then it must satisfy that
\begin{align}
dist(p,p')\leq dist(p,p^\times).
    \label{eq:lemma2_1}
\end{align}
According to the triangle inequality, we can obtain
\begin{align}
dist(p',p^+)\leq dist(p,p')+dist(p,p^+).
    \label{eq:lemma2_2}
\end{align}
Combining Inequality~\eqref{eq:lemma2_1} and Inequality~\eqref{eq:lemma2_2}, we can obtain
\begin{align}
dist(p',p^+)&\leq dist(p,p^\times)+dist(p,p^+)\nonumber\\
&=dist(p^\times,p^+)\nonumber\\
&=r_{p^+}.
    \label{eq:lemma2_3}
\end{align}
From above, we can construct a corollary that any point lying in $C_{p:dist(p,p^\times)}$ must belong to $k$NN($p^+$). 
Specifically, the number of points lying in $C_{p:dist(p,p^\times)}$ must not be greater than $k$, i.e., the size of $k$NN($p^+$).
Equivalently, there is no more than $k$ points closer to $p$ than $p^\times$.
Thus, $p_k$ (the $k$th closest point to $p$) cannot be closer than $p^\times$ to $p$.
Then $dist(p,p^\times) \leq dist(p,p_k)=r_p$.
Suppose Inequality~\eqref{eq:lemma2_0} holds,
\begin{align}
dist(p,q)& \leq r_{p^+}-dist(p,p^+)\nonumber\\
&=dist(p^\times,p^+)-dist(p,p^+)\nonumber\\
&=dist(p,p^\times)\leq r_p.
    \label{eq:lemma2_4}
\end{align}
From Lemma~\ref{lemma_1} and Inequality~\eqref{eq:lemma2_4}, we can deduce that $p \in$ R$k$NN($q$).
Therefore Lemma~\ref{lemma_2} proved to be true.
\end{proof}

Lemma~\ref{lemma_2} provides a sufficient but unnecessary condition for determining that a point belongs to R$k$NN($q$), where $q$ represents the query point. That means if a point $p$ satisfies the condition of Inequality~\eqref{eq:lemma2_0}, it can be determined as one of R$k$NN($q$) without issuing a $k$NN query.
In the case that $r_{p^+}$ is known, we can verify whether Inequality~\eqref{eq:lemma2_0} holds by only calculating the Euclidean distance from $p$ to $q$ and $p^+$ respectively.
Calculating the Euclidean distance between two points can be regarded as an atomic operation. Hence the computational complexity of the verification method corresponding to Lemma~\ref{lemma_2} is $O(1)$.
\begin{figure}[htbp]
    \centering
    \includegraphics[width=0.46\textwidth]{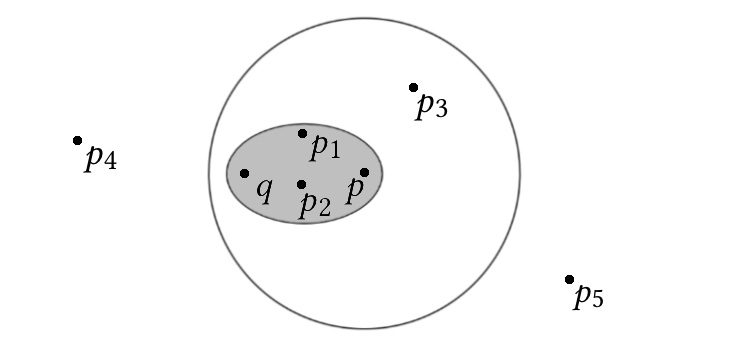}
    \caption{Positive determine region}
    \label{fig:positive_discriminant_region}
\end{figure}

\begin{Definition}
\textbf{Positive determine region:}
Given the query point $q$ and a point $p$, the positive determine region of $p$ is the internal region of $E_{p:q}^{r_p}$. Formally, it is denoted as $RG_{det}^+(p)$ and is defined as follows:
\begin{align}
RG_{disc}^+(p) = \{p'\;|\;dist(p',q)+dist(p',p) \leq r_p\}.
    \label{eq:positive_discriminant_region}
\end{align}
\end{Definition}

From the triangle inequality, it can be shown that
\begin{align}
dist(p',q)+dist(p',p) \geq dist(p,q).
\end{align}
If $p\notin$ R$k$NN($q$), i.e., $dist(p,q)>r_p$, 
\begin{align}
dist(p',q)+dist(p',p) > r_p
\end{align}
then $RG_{det}^+(p)=\emptyset$. Therefore, if  $RG_{det}^+(p)\neq\emptyset$, $p$ must belong to R$k$NN($q$). In consequence, from Lemma~\ref{lemma_2}, we can construct a corollary that, for any point $p$, if $RG_{det}^+(p)$ is not empty, all the points lying inside of $RG_{det}^+(p)$ must belong to R$k$NN($q$). 

As shown in Figure~\ref{fig:positive_discriminant_region}, $q$ represents the query point, the internal region of the circle $C_{p:r_p}$ indicates $RG${\tiny$_{k{\rm NN}}$}$(p)$, and the gray region within the ellipse $E_{p:q}^{r_p}$ is for $RG_{det}^+(p)$.
As $p_1$ and $p_2$ lies in $RG_{det}^+(p)$, we can know $p_1,p_2\in$ R$k$NN($q$). Whereas $p_3$, $p_4$ and $p_5$ lie out of $RG_{det}^+(p)$, so we cannot directly determine whether or not they belong to R$k$NN($q$) by Lemma~\ref{lemma_2}.

\begin{Lemma}
\label{lemma_3}
Given a query point $q$ and a point $p^- \notin$ R$k$NN($q$), a point $p$ cannot be any one of R$k$NN($q$) if it satisfies
\begin{align}
\label{eq:lemma3_0}
dist(p,q)-dist(p,p^-)>r_{p^-}.
\end{align}
\end{Lemma}

\begin{figure}[htbp]
    \centering
    \includegraphics[width=0.25\textwidth]{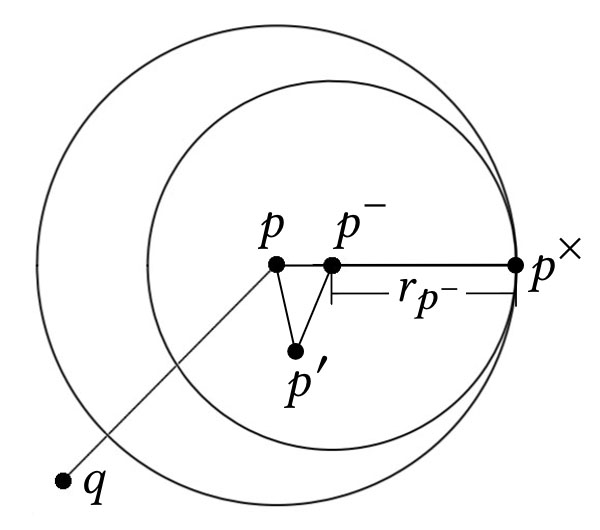}
    \caption{Lemma~\ref{lemma_3}}
    \label{fig:lemma3}
\end{figure}

\begin{proof}
As shown in Figure~\ref{fig:lemma3}, the smaller circle takes $p^-$ as the center and $r_{p^-}$ as the radius, which represents the $k$NN region of $p^-$.
The point $p^\times$ is the intersection of an extension of $L_{p,p^-}$ (a line segment between $p$ and $p^-$) with $C_{p^-:r_{p^-}}$.
The larger circle takes $p$ as the center and $dist(p,p^\times)$ as the radius.
Let $p'$ be an arbitrary point inside of $C_{p^-:r_p-}$, then it must satisfy that
\begin{align}
dist(p^-,p')\leq dist(p^-,p^\times) = dist(p^-,p^\times).
    \label{eq:lemma3_1}
\end{align}
According to the triangle inequality, we can obtain
\begin{align}
dist(p,p')\leq dist(p,p^-)+dist(p^-,p').
    \label{eq:lemma3_2}
\end{align}
From Inequality.\eqref{eq:lemma3_1} and Inequality.\eqref{eq:lemma3_2}, we can get that
\begin{align}
dist(p,p')&\leq dist(p,p^-)+dist(p^-,p^\times)\nonumber\\
&=dist(p,p^\times)\nonumber\\
&=r_p.
    \label{eq:lemma3_3}
\end{align}
Then we realize that all the points lying in $RG${\tiny$_{k{\rm NN}}$}$(p^-)$ must lie inside $C_{p:dist(p,p^\times)}$, namely the number of points lying inside of $C_{p:dist(p,p^\times)}$ must be no less than $k$, i.e., the number of points lying in $RG${\tiny$_{k{\rm NN}}$}$(p^-)$.
That is to say, there exist at least $k$ points no further than $p^\times$ away from $p$. Equivalently, $dist(p,p^\times) \geq dist(p,p_k)=r_p$ (where $p_k$ represents the $k$th closest point to $p$).
If the condition of Inequality~\eqref{eq:lemma3_0} is satisfied,
\begin{align}
dist(p,q)& > dist(p,p^-)+r_{p^-}\nonumber\\
&=dist(p,p^-)+dist(p^-,p^\times)\nonumber\\
&=dist(p,p^\times)\geq r_p.
    \label{eq:lemma3_4}
\end{align}
From Lemma~\ref{lemma_1} and Inequality~\eqref{eq:lemma3_4}, we can deduce that $p \notin $ R$k$NN($q$).
Therefore, Lemma~\ref{lemma_3} proved to be true.
\end{proof}

From Lemma~\ref{lemma_3}, we can know that, if a point is determined not to be one of R$k$NN($q$) and its $k$NN radius is known, then there may exist some other points that can be sufficiently determined to belong to R$k$NN($q$) without performing a $k$NN query but by performing two times of simple Euclidean distance calculation. That means the computational complexity of the verification method based on Lemma~\ref{lemma_3} is $O(1)$.

\begin{figure}[htbp]
    \centering
    \includegraphics[width=0.45\textwidth]{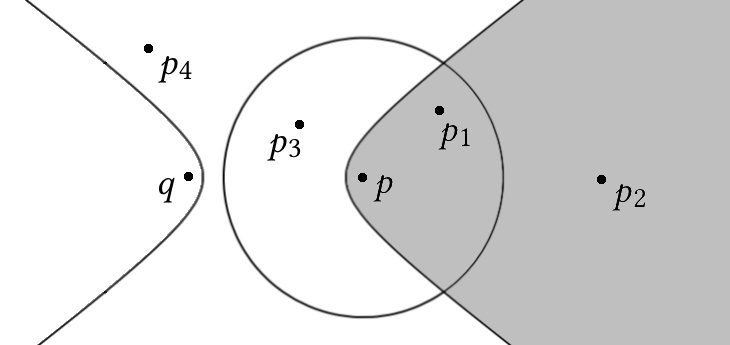}
    \caption{Negative determine region}
    \label{fig:negative_discriminant_region}
\end{figure}

\begin{Definition}
\textbf{Negative determine region:}
Given the query point $q$ and a point $p$, $H_{p:q}^{r_p}$ divides the space into three regions of which the one contains $p$ is the negative determine region of $p$. Formally, this region is denoted as $RG_{det}^-(p)$ and is defined as follows:
\begin{align}
RG_{det}^-(p) = \{p'\;|\;dist(p',q)-dist(p',p) > r_p\}.
    \label{eq:negative_discriminant_region}
\end{align}
\end{Definition}

For an arbitrary point $p'$,from the triangle inequality in $\bigtriangleup pqp'$, it can be known that
\begin{align}
dist(p',p)+dist(p,q) \geq dist(p',q).
\end{align}
If $p\in$ R$k$NN($q$), i.e., $dist(p,q)\leq r_p$, 
\begin{align}
dist(p',q)-dist(p',p) \leq dist(p,q)\leq r_p
\end{align}
then $RG_{det}^-(p)=\emptyset$. Therefore, if  $RG_{det}^-(p)$ is not empty, $p$ must belong to R$k$NN($q$). Hence from Lemma~\ref{lemma_3}, we can draw such a corollary that, for an arbitrary point $p$, if $RG_{det}^-(p)$ is not empty, any point lying inside $RG_{det}^-(p)$ cannot belong to R$k$NN($q$). 

As shown in Figure~\ref{fig:negative_discriminant_region}, $q$ represents the query point, the region within the circle centered on $p$ represents $RG${\tiny$_{k{\rm NN}}$}$(p)$, and the gray region separated by the hyperbola $H_{p:q}^{r_p}$ on the right represents $RG_{det}^-(p)$. 
As in the figure, $p_1$ and $p_2$ lie inside $RG_{det}^-(p)$, while $p_3$ and $p_4$ do not.
Then we can determine that $p_1$ and $p_2$ must not belong to R$k$NN($q$), whereas we cannot tell by Lemma~\ref{lemma_3} whether $p_3$ or $p_4$ belongs to R$k$NN($q$) or not.

\begin{Definition}
\textbf{Positive/Negative determine point:}
Given the query point $q$ and two other points $p$ and $p'$, if $p'$ lies in $RG_{det}^+(p)$, we claim that $p$ is a positive determine point of $p'$ and $p$ can positive determine $p'$. It is denoted as $p\xrightarrow{+det}p'$.
Similarity, if $p'$ lies in $RG_{det}^-(p)$, we name that $p$ is a negative determine point of $p'$ and $p$ can negative determine $p'$. It is denoted as $p\xrightarrow{-det}p'$.
If not specified, both of these two types of points may be collectively referred to as determine points and we can use $p\xrightarrow{det}p'$ to express that $p$ can dedermine $p'$.
\end{Definition}

Whether a point belongs to the R$k$NN set of the query point or not, the corresponding verification method with low computational complexity is provided.
However, when performing the verification of Lemma~\ref{lemma_2} or Lemma~\ref{lemma_3}, the distance from the point to be determined to the query point and the positive/negative determine point should be calculated respectively.
In order to further improve the verification efficiency of some points, we propose Lemma~\ref{lemma_4}.

\begin{Lemma}
\label{lemma_4}
Given a query point $q$, a point $p$ must be one of R$k$NN($q$) if it satisfies
\begin{displaymath}
dist(p,q)\leq r_{q}/2.
\end{displaymath}
\end{Lemma}

\begin{figure}[htbp]
    \centering
    \includegraphics[width=0.25\textwidth]{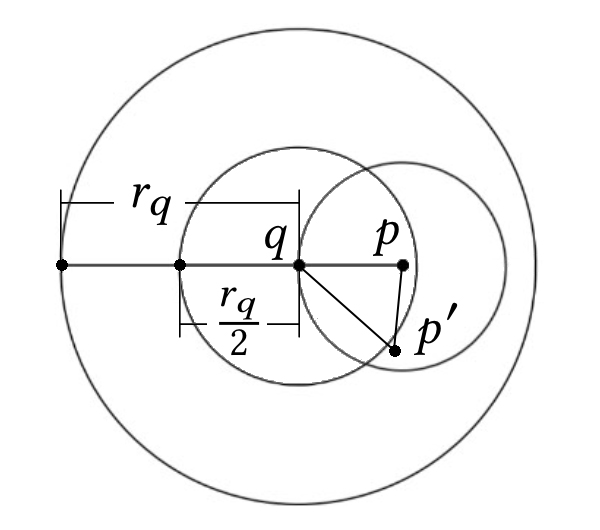}
    \caption{Lemma~\ref{lemma_4}}
    \label{fig:lemma4}
\end{figure}

\begin{proof}
In Figure~\ref{fig:lemma4}, there are three circles, two of which are centered on $q$ and take $r_q$ and $r_q/2$ as the length of their radii, respectively.
The other circle takes $p$ as the center and $dist(p,q)$ as the length of the radius, where $p$ lies in $c_{q:r_q/2}$, i.e., $dist(q,p)\leq r_q/2$.
Let $p'$ be an arbitrary point inside of $C_{p:dist(p,q)}$, then it must satisfy that
\begin{align}
    dist(p,p') \leq dist(q,p).
\end{align}
From the triangle inequality of $\bigtriangleup pqp'$, it can be obtained that
\begin{align}
    dist(q,p')\leq dist(q,p)+dist(p,p').
\end{align}
Then we can get that,
\begin{align}
    dist(q,p')\leq 2\cdot dist(q,p).
\end{align}
Because $dist(q,p)\leq r_q/2$,
\begin{align}
    dist(q,p')\leq 2\cdot r_q/2=r_q
\end{align}
That means, any point lying in $C_{p:dist(p,q)}$ must belong to $k$NN($q$). Therefore, the number of points lying in $C_{p:dist(p,q)}$ must not be greater than $k$, i.e., the size of $k$NN($q$), which means there is no more than $k$ points closer to $p$ than $q$.
Hence $p_k$ ($k$th closest point to $p$) cannot be closer than $q$ to $p$.
Then
\begin{align}
   dist(p,q) \leq dist(p,p_k)=r_p.
\end{align}
According to Lemma~\ref{lemma_1}, $p\in$ R$k$NN($q$), then Lemma~\ref{lemma_4} is proved.
\end{proof}

\begin{figure}[htbp]
    \centering
    \includegraphics[width=0.45\textwidth]{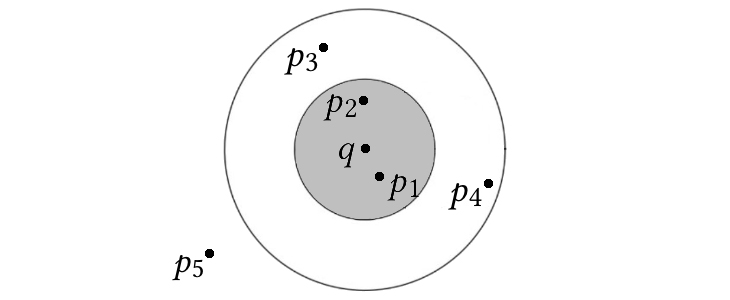}
    \caption{Semi-$k$NN region}
    \label{fig:semi_knn_region}
\end{figure}
\begin{Definition}
\textbf{Semi-$k$NN region:}
Given the query point $q$, the semi-$k$NN region of $q$ is the internal region of $C_{q:r_q/2}$. Formally, it is denoted as $SRG${\tiny$_{k{\rm NN}}$}$(q)$ and is defined as Equation~\eqref{eq:semi_knn_region}.
\begin{align}
SRG_{k{\rm NN}}(q) = \{p\;|\;dist(p,q) \leq r_q/2\}
    \label{eq:semi_knn_region}
\end{align}
\end{Definition}

As shown in Figure~\ref{fig:semi_knn_region}, $q$ represents the query point, the region within the larger circle represents $RG${\tiny$_{k{\rm NN}}$}$(q)$, and the gray region within the smaller circle represents $SRG${\tiny$_{k{\rm NN}}$}$(q)$. 
It can be observed from the figure, $p_1$ and $p_2$ lie in the gray region, while $p_3$, $p_4$ and $p_5$ do not.
Then $p_1$ and $p_2$ can be determined as members of R$k$NN($q$). Nevertheless, we cannot determine whether $p_3$, $p_4$ or $p_5$ belongs to R$k$NN($q$) or not by Lemma~\ref{lemma_4}

With Lemma~\ref{lemma_1}, \ref{lemma_2}, \ref{lemma_3} and \ref{lemma_4}, we can find all the points in the R$k$NNs of the query point by verifying only a small portion of points in the candidates.

\subsection{Selection of determine points}
Theoretically, when using Lemma~\ref{lemma_1}, \ref{lemma_2}, \ref{lemma_3} and \ref{lemma_4} to verify the candidates, any R$k$NN point can be considered as a positive determine point. 
Similarly, if a point is not a member of R$k$NNs, then it can be considered as a negative determine point.
In other words, all points in the candidate set are eligible to be selected as determine points.
Our aim is to issue as few $k$NN queries as possible in the process of R$k$NN queries, that is, to use as few determine points as possible to determine all the other points in the candidate set.
Therefore, the selection of determine points is very important for improving the efficiency of R$k$NN queries.
Which points should be selected as  determine points is what we will scrutinize next.
\begin{Definition}
\textbf{Determine point set:} For a R$k$NN query, given a set $S_{cnd}$ of candidates and denoted as $S_{dist}$, a determine set is such a set that the following condition is satisfied:
\begin{align}
    \forall p\in S_{cnd}\setminus S_{dist},\; \exists p'\in S_{dist}: p'\xrightarrow{det}p .
\end{align}

\end{Definition}
Because it is not certain how many points and which points need to be selected as determine points, the total number of schemes for selecting  determine points can be as large as $\sum\limits_{i=1}^{|S_{cnd}|}\dbinom{|S_{cnd}|}{i}$, where $|S_{cnd}|$ means the number of candidates. Hence the computational complexity of finding the absolute optimal one out of all the schemes is as much as $O(k!)$.
However, it is not difficult to come up with a relatively good determine points selecting scheme, of which the size of the determine set $|S_{dist}|$ is just about $O(\sqrt{k})$.

For a positive determine point, most of the points in its determine region are closer to the query point than itself.
Furthermore, any negative determine point is closer to the query point than most of the points in its own determine region.
Therefore, a point belonging to R$k$NNs can rarely be determined by a point closer to the query point than itself, and the probability that a point not belonging to R$k$NNs can be determined by a point further than itself away from the query point is also very low.
Therefore, the points which are extremely close to the boundary of the R$k$NN region (i.e., influence zone \cite{InfZone}) are rarely able to be determined by other points.
Thus, these points should be selected as determine points in preference.
However, it is impossible to directly find these points near the boundary without pre-calculating the R$k$NN region.
Calculating the R$k$NN region is a very computational costly process for its computational complexity of $O(k^3)$.
While the $k$NN region of the query point is easy to obtained by issuing a $k$NN query.
Assuming that the points are uniformly distributed, the $k$NN region and the R$k$NN region of a query point are extremely approximate and the difference between them is negligible.
Hence it is a good strategy to preferentially select the points near the boundary of $k$NN region as the determine points to some extent.

\begin{figure}[htbp]
    \centering
    \includegraphics[width=0.3\textwidth]{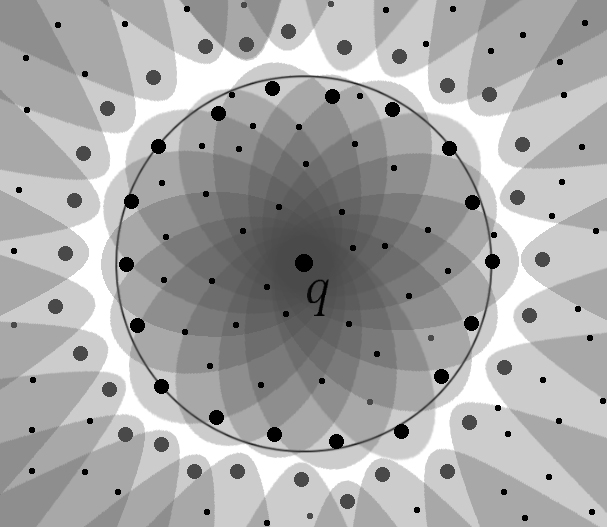}
    \caption{Determine point set}
    \label{fig:discriminant_set}
\end{figure}
As shown in Figure~\ref{fig:discriminant_set}, there are some points distributed.
The region inside the circle with $q$ as the center represents the $k$NN region of $q$.
In general, only the points near the boundary of $RG${\tiny$_{k{\rm NN}}$}$(q)$ need to be selected as the determine points and all the other candidate points can be determined by these determine points.
In other words, if the points are evenly distributed, the points near the boundary of $RG${\tiny$_{k{\rm NN}}$}$(q)$ are enough to form a valid determine set of $q$.
Because the distribution of points is not guaranteed to be absolute uniform, it is not always reliable if only the points near the boundary of the kNN region of the query point are taken as determine points for a R$k$NN query.

In order to ensure the reliability of the selection, we propose a strategy to dynamically construct the determine set while verifying the candidate points.
First, the candidate points  belonging to $k$NN($q$) are accessed in descending order of distance to $q$. Then the other candidate points are accessed in ascending order of distance to $q$.
During the process of accessing  candidates,
once the currently accessed point cannot be determined by any point in the determine point set, this point should be selected as a determine point and put into the determine point set. Otherwise, we can use a corresponding point in the determine point set to determine whether it belongs to R$k$NNs or not.
\subsection{Matching~candidate~points~with~determine~points}
Under the above strategy, it is sufficient to ensure that any point  not belonging to $S_{dist}$ can be determined by at least one point in $S_{dist}$.
Since the expected size of $S_{dist}$ is $O(\sqrt{k})$ (see Section~5),
the computational complexity  of finding a determine point for a point by exhaustive searching the determine set is $O(\sqrt{k})$.
Obviously, it is not a good idea to match candidate points with their determine point in this way.
Therefore, we propose a method based on Voronoi diagrams to improve the efficiency of this process.

Given a Voronoi diagram $VD(P)$ of a point set $P$ and a continuous region $RG$, the vast majority of points in $RG$ have at least one Voronoi neighbor lying in $RG$ \cite{DBLP:journals/corr/abs-1912-00426}.
For any determine point, its determine region is a continuous region (ellipse region or hyperbola region).
So for a non-determine point, there is high probability that at least one of its Voronoi neighbors can determine it or shares a determine point with it.
Therefore, when accessing a candidate point, if the point can be determined by one of its Voronoi neighbors or the determine point of one of its Voronoi neighbors, this point can be determined whether belongs to the R$k$NNs. Otherwise, we say that this point is almost impossible to be determined by any known determine point and it should be marked as a determine point.
Recall Lemma~\ref{lemma:not_exceed_6}, in two dimensions, the expected number of Voronoi neighbors per point is 6, which is a constant.
By using the above approach we can find the determine point for a non-determine point with a computational complexity of $O(1)$.
\subsection{Algorithm}
In this subsection, we will introduce the implementation of the R$k$NN algorithm based the above approaches.

The pseudocode for the verification methood is shown in Algorithm~\ref{alg:csd}.
When verifying a point, we first try to determine whether the point belongs to R$k$NNs by Lemma 4 (line 2).
If this fails, we visit the Voronoi neigbors of the point and try to use Lemma 2 or Lemma3 to determine it (line 10 and line 13).
If none of the three lemmas above apply to this point, then we issue a $k$NN query for it and use Lemma 4 to verify it (line 18).


\begin{algorithm}[htbp]
\caption{\textbf{verify(}$p, q, k, r_q, S_v, S_{det}, D_{det}$\textbf{)}}
\label{alg:csd}
\KwIn{the point $p$ to be verified, the query point $q$, the parameter $k$, the $k$NN radius $r_q$ of $q$, the set $S_v$ of points that have been visited , the determine point set $S_{det}$ and the dictionary $D_{det}$ that records the corresponding determine points for non-determine points}
\KwOut{whether $p \in$ R$k$NN$(q)$.}
$S_v$.add($p$)\;
\If(\tcc*[f]{Lemma~\ref{lemma_4}}){$dist(p,q)\leq r_q/2$}
{
    \Return \True\;
}
\ForEach{$p_n \in$ VN($p$)}
{
    \If{$p_n \in S_v$}
    {
        \eIf{$p_n \in S_{det}$}
        {
            $p_{det} \longleftarrow p_n$\;
        }
        {
            $p_{det} \longleftarrow D_{det}[p_n]$\;
        }
        
        \If(\tcc*[f]{Lemma~\ref{lemma_2}}){$p_{det} \in$~ R$k$NN$(q)$~\And~$dist(p,q)+dist(p,p_{det}) \leq r_{p_{det}}$}
        {
            $D_{det}[p] \longleftarrow p_{det}$\;
            \Return \True\;
        }
        \If(\tcc*[f]{Lemma~\ref{lemma_3}}){$p_{det} \notin$~R$k$NN$(q)$~\And~$dist(p,q)-dist(p,p_{det}) > r_{p_{det}}$}
        {
            $D_{det}[p] \longleftarrow p_{det}$\;
            \Return \False\; 
        }
    }
}
$r_p \longleftarrow$ calculate the $k$NN radius of $p$\;
$S_{det}$.add($p$)\;
\eIf(\tcc*[f]{Lemma~\ref{lemma_1}}){$r_p \geq dist(p,q)$}
{
    \Return \True\;
}
{
    \Return \False\; 
}
\end{algorithm}

\begin{algorithm}[htbp]
\caption{\textbf{R$k$NN(}$q$\textbf{)}}
\label{alg:rknn}
\KwIn{the query point $q$}
\KwOut{R$k$NN($q$}
$S_{cnd} \longleftarrow$ \textbf{generateCandidates}($q,k$)\;
Sort $S_{cnd}$ in ascending order by the distance to $q$\;
$r_q \longleftarrow$ calculate the $k$NN radius of $q$\;
$S_v \longleftarrow \emptyset$\;
$S_{det} \longleftarrow \emptyset$\;
$D_{det} \longleftarrow$ generate an empty dictionary\;
$S_{{\rm R}k{\rm NN}} \longleftarrow \emptyset$\;
\For{$i \longleftarrow k$ \KwTo $1$}
{
    \If{\textbf{verify}($S_{cnd}[i],q,k,r_q,S_{v},S_{det},D_{det}$)}
    {
        $S_{{\rm R}k{\rm NN}}.add(S_{cnd}[i])$\;
    }
}
\For{$i \longleftarrow k+1$ \KwTo $6k$}
{
    \If{\textbf{verify}($S_{cnd}[i],q,k,r_q,S_{v},S_{det},D_{det}$)}
    {
        $S_{{\rm R}k{\rm NN}}.add(S_{cnd}[i])$\;
    }
}
\Return $S_{{\rm R}k{\rm NN}}$\;
\end{algorithm}

Using the verification approach in Algorithm~\ref{alg:csd}, we implement an efficient R$k$NN algorithm, as shown in Algorithm~\ref{alg:rknn}.
First we generate the candidate set in the same way as VR-R$k$NN \cite{VoR}, where the size of candidate is 6$k$ (line 1). 
Next, the candidate set is sorted in ascending order by the distance to the query point (line 2).
Then the first $k$ elements of the candidate set and the rest of the elements are divided into two groups.
The elements in the two groups are verified one by one in the order from back to front and from front to back, respectively (line 8 and line 11).
After all candidate points are verified, the R$k$NNs of the query point is obtained.

\begin{algorithm}[htbp]
\caption{\textbf{pruning(}$q, k$\textbf{)}}
\label{alg:generate_candidates}
\KwIn{the query point $q$ and the parameter $k$}
\KwOut{the candidates of R$k$NN($q$)}
$H \longleftarrow MinHeap()$\;
$Visited \longleftarrow \emptyset$\;
\For{$i \longleftarrow 1$ \KwTo $6$}
{
    $S_{cnd}[i] \longleftarrow MinHeap()$\;
}
\ForEach{$p \in$ VN($q$)}
{
    $H.push([1,p])$\;
    $Visited.add(p)$\;
}
\While{$|H|>0$}
{
    $[dist_{DG}(p), p] \longleftarrow H.pop()$\; 
    \For{$i \longleftarrow 1$ \textbf{to} $6$}
    {
        \If{$Segment_i$ contains $p$}
        {
            \eIf{$|S_{cnd}[i]| >0$}
            {
                $p_n \longleftarrow$ the last point in $S_{cnd}[i]$\;
            }
            {
                $p_n \longleftarrow$ a point infinitely away from $q$\;
            }
            
            \If{$dist_{DG}(p) \leq k$ \And $dist(q,p)\leq dist(q,p_n)$}
            {
                $S_{cnd}[i].push([dist(p,q),p])$\;
                \ForEach{$p' \in$ VN($p$)}
                {
                    \If{$p' \notin Visited$}
                    {
                        $dist_{DG}(p') \longleftarrow dist_{DG}(p)+1$\;  $H.push([dist_{DG}(p'),p'])$\;
                        $Visited.add(p')$\;
                    }
                }
            }
        }
    }
}
$Candidates \longleftarrow \emptyset$\;
\For{$i \longleftarrow 1$ \textbf{to} $6$}
{
    \For{$j \longleftarrow 1$ \textbf{to} $k$}
    {
        $Candidates$.add($S_{cnd}[i]$.pop())\;
    }
    
}
\Return $Candidates$;
\end{algorithm}
We used the same algorithm as VR-R$k$NN to generate the candidate set, and we do not improve it.
The core of this algorithm is still from the Six-regions \cite{six}.
In addition, it uses a Voronoi diagram to find the candidate points incrementally according to Lemma~\ref{lemma:nn}.
By Lemma~\ref{lemma:dgd},  only the points whose Delaunay distance to the query point is not larger than $k$ are eligible to be selected as candidate points.
Hence the number of points accessed  for finding candidates in the algorithm is guaranteed to be no more than $O(k^2)$.
The pseudocode of the algorithm for generating candidates is presented in Algorithm~\ref{alg:generate_candidates}.

\section{Theoretical analysis}
In this section, we analyze the expected size of determine point set, the expected number of accessed points and the computational complexity of our algorithm.
\subsection{Expected size of determine point set}
\label{discriminant_set_analysis}
The query point is $q$, the number of points in R$k$NN($q$) is $|$R$k$NN$|$, and the number of points near the boundary of $RG${\tiny$_{k{\rm NN}}$}$(q)$ is $|S_b|$.
The area and circumference (total length of the boundary) of $RG${\tiny$_{k{\rm NN}}$}$(q)$ are denoted as $A${\tiny$_{{\rm R}k{\rm NN}}$}$(q)$ and $C${\tiny$_{{\rm R}k{\rm NN}}$}$(q)$, respectively.
The expected size of the determine point set of $q$ is $|S_{det}|$.

It is shown that the expected value of $|$R$k$NN$|$ is $k$ \cite{InfZone}.
Thus, the radius of the approximate circle of $RG${\tiny$_{k{\rm NN}}$}$(q)$ is equal to $r_q$. Then
\begin{align}
\label{area}
A_{_{{\rm R}k{\rm NN}}}(q)=\pi\cdot {r_q}^2 
\end{align}
\begin{align}
\label{circumference}
C_{_{{\rm R}k{\rm NN}}}(q)=2\pi\cdot r_q.
\end{align}
The following equation can be obtained from Equation~\eqref{area} and Equation ~\eqref{circumference}.
\begin{align}
C_{_{{\rm R}k{\rm NN}}}(q) =2\sqrt{\pi\cdot A_{_{{\rm R}k{\rm NN}}}(q)}
\end{align}
As the points around the boundary of $RG${\tiny$_{k{\rm NN}}$}$(q)$ consists of two sets of points where one is inside $RG${\tiny$_{k{\rm NN}}$}$(q)$ and the other is outside,
$|S_{b}|$ is to $|$R$k$NN$|$ what $2\cdot C${\tiny$_{{\rm R}k{\rm NN}}$}$(q)$ is to $A${\tiny$_{{\rm R}k{\rm NN}}$}$(q)$, i.e.,
\begin{align}
|S_{b}|= 2\cdot 2\sqrt{\pi\cdot |{\rm R}k{\rm NN}|}=4\sqrt{\pi\cdot k} \approx 7.1\sqrt{k}.
\end{align}
If all the points near the boundary are selected as the determine points, there must be some redundancy, i.e., the determine region of some points will overlap.
Hence the size of the determine point set generated under our strategy is less than the number of the points near the boundary of the R$k$NN region, i.e, $|S_{det}|\leq 7.1\sqrt{k}$.

\subsection{Expected number of accessed points}
\label{accessed_poing_analysis}
For an R$k$NN query of $q$, the candidate points are distributed in an approximately circular region $RG$$_{cnd}$$(q)$ centered around $q$, which has an area $A_{cnd}(q)$ and  a circumference $C_{cnd}(q)$.
The expected number of accessed points is $|S_{ac}|$.
In the filtering phase of our approach, the points accessed include all the the candidate points and their Voronoi neighbors.
Except for the points in the candidate set, the other accessed points are distributed outside $RG$$_{cnd}$$(q)$ and adjacent to the boundary of $RG$$_{cnd}$$(q)$. Hence $|S_{ac}|-|S_{cnd}|$ is to $|S_{cnd}|$ what $C_{cnd}(q)$ is to $A_{cnd}(q)$, i.e.,
\begin{align}
|S_{ac}|-|S_{cnd}|&= 2\sqrt{\pi\cdot |S_{cnd}|}
\end{align}
\begin{align}
|S_{ac}|&= |S_{cnd}|+2\sqrt{\pi\cdot |S_{cnd}|}\nonumber\\
&=6k+2\sqrt{\pi\cdot 6k}\approx 6k+8.7\sqrt{k}
\end{align}
Therefore, if the points are distributed uniformly, the expected number of accessed points is approximately $6k+8.7\sqrt{k}$.
When the points are distributed unevenly, $|S_{ac}|$ becomes larger. However, it has an upper bound.
Recall Lemma~\ref{lemma:dgd}, we can make deduce that
only the points whose Delaunay graph distance to $q$ is not larger than $k$ are eligible to be selected as candidate points. Then
\begin{align}
|S_{ac}|\leq \sum\limits^k_{i=1}2\pi\cdot i=(k^2+k)\pi .
\end{align}

\subsection{Computational complexity}
The expected computational complexity of the filtering phase of our approach is $O(k\cdot log\,k)$ \cite{VoR}.
In the refining phase, we have to issue a $k$NN query with $O(k\cdot log\,k)$ computational complexity for each determine point, and the size of the determine point set is about $7.1\sqrt{k}$. The other candidates only need to be verified by our efficient verification method. Thus, the computational complexity of the refining phase is $O(k^{1.5}\cdot log\,k)$.
Hence the overall computational complexity of our R$k$NN algorithm is $O(k^{1.5}\cdot log\,k)$.

\section{Experiments}
In the previous section, we discussed the theoretical performance of our algorithm. In this section, we intend to evaluate the performance of aspects through comparison experiments.

\subsection{Experimental settings}
In the experiments, we let VR-R$k$NN \cite{VoR}  and the state-of-the-art R$k$NN approach SLICE \cite{SLICE} to be the competitors of our method.

The settings of our experiment environment are as follows. The experiment is conducted on a personal computer with Python 2.7. The CPU is Intel Core i5-4308U 2.80GHz and the RAM is DDR3 8G.

To be fair, all three methods in the experiment are implemented in Python, with six partitions in the pruning phase.
We use two types of experimental data sets: simulated data set and real data set\footnote{49,601 non-duplicative data points on the geographic coordinates of the National Register of Historic Places (\url{http://www.math.uwaterloo.ca/tsp/us/files/us50000_latlong.txt})}.
To decrease the error of the experiments, we repeat each experiment for 30 times
and calculate the average of the results. The query point for each time of the experiment is randomly generated.

Our experiments are designed into four sets. 
The first set of experiments is used to evaluate the effect of the data size on the time cost of the R$k$NN algorithms. The data size
is from $10^3$ to $10^6$ and the value of $k$ is fixed at 200.
The rest of sets are used to evaluate the effect of the value of $k$ on the time cost, the number of verified points and the number of the accessed points of the R$k$NN algorithms, respectively. For these three sets of experiments, the size of the simulated data is fixed at $10^6$, the size of the real data is 49,601 and the value of $k$ varies from $10^1$ to $10^4$.

\subsection{Experimental results}
\begin{table}[htbp]
\centering
\caption{Total time cost(in ms) of different R$k$NN algorithms with various sizes of data sets.}
\label{table:data_size}
\begin{tabular}{@{}crrrr@{}}
\toprule
\multirow{2}{*}{\textbf{Algorithm}} & \multicolumn{4}{c}{\textbf{Data size}}\\ \cmidrule(l){2-5}
& \textbf{\quad\quad $10^3$} & \textbf{\quad\quad $10^4$}  & \textbf{\quad\quad $10^5$}   & \textbf{\quad\quad $10^6$}   \\ \midrule
VR-R$k$NN  & 510      & 725       & 728      & 732\\
SLICE  & 232      & 397      & 438     & 441\\
Our approach  & \textbf{59}      & \textbf{65}      & \textbf{69}     & \textbf{72}\\
\bottomrule
\end{tabular}
\end{table}
\begin{figure}[htbp]
    \centering
    \includegraphics[width=0.32\textwidth]{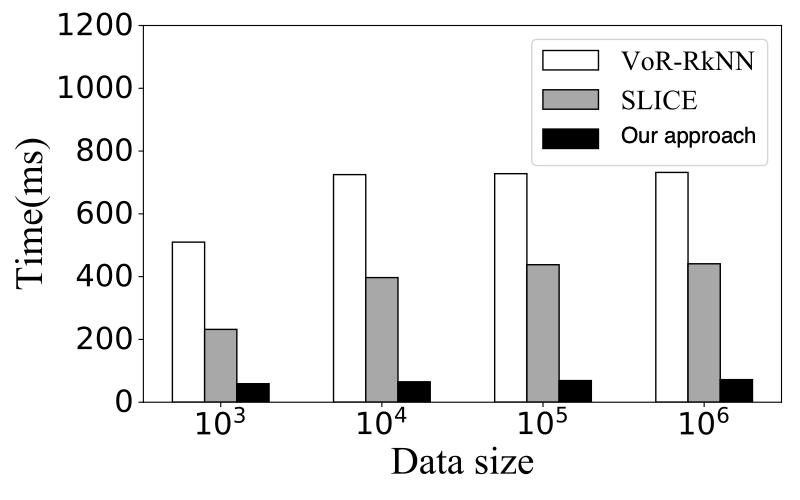}
    \caption{Effect of data size on efficiency of R$k$NN queries}
    \label{fig:time_size}
\end{figure}
Figure~\ref{fig:time_size} shows the time cost of the three R$k$NN algorithms with various data sizes.
As we can see, when the number of points in the database is significantly much lager than $k$, the impact of the data size on the time cost of R$k$NN queries is very limited.
If the number of points in the database is small enough to be on the same order of magnitude as $k$,
all points in the database become candidate points. Then the smaller the database size, the less time cost of the R$k$NN query.
When the number of points in the database is above 10,000 and the value of $k$ is fixed at 200, the time cost of our approach is always around 84\% and 90\% less than that of SLICE and VR-R$k$NN, respectively.
The detailed experimental results are presented in Table~\ref{table:data_size}.

\begin{table}[htbp]

\caption{Total time cost (in ms) of R$k$NN queries with various values of $k$.}
\label{table:time_k}
\centering
\begin{tabular}{@{}crrrrrr@{}}
\toprule
\multirow{2}{*}{\textbf{$k$}} & \multicolumn{3}{c}{\small \textbf{Simulated data}} & \multicolumn{3}{c}{\small \textbf{Real data}} \\ \cmidrule(lr){2-4} \cmidrule(l){5-7}
& {\fontsize{5.5}{0}\textbf {VR-R$k$NN}} & {\fontsize{5.5}{0}\textbf{SLICE}}  & {\fontsize{5.5}{0}\textbf{Our approach}}   & {\fontsize{5.5}{0}\textbf{VR-R$k$NN}}   & {\fontsize{5.5}{0}\textbf{SLICE}} & {\fontsize{5.5}{0}\textbf{Our approach}}\\ \midrule
$10^1$  & 5      & 26       & \textbf{2}      & 4      & 28      & \textbf{2}\\
$10^2$  & 199      & 193      & \textbf{39}     & 194     & 283      & \textbf{29}\\
$10^3$  & 20576      & 3759      & \textbf{801}     & 17212     & 4610      & \textbf{813}\\
$10^4$  & 2118391      & 321233      & \textbf{22077}     & 1829742     & 226959      & \textbf{23911}\\
\bottomrule
\end{tabular}
\end{table}
\begin{figure}[htbp]
  \centering
  \subfigure[Simulated data]{\includegraphics[width=0.235\textwidth]{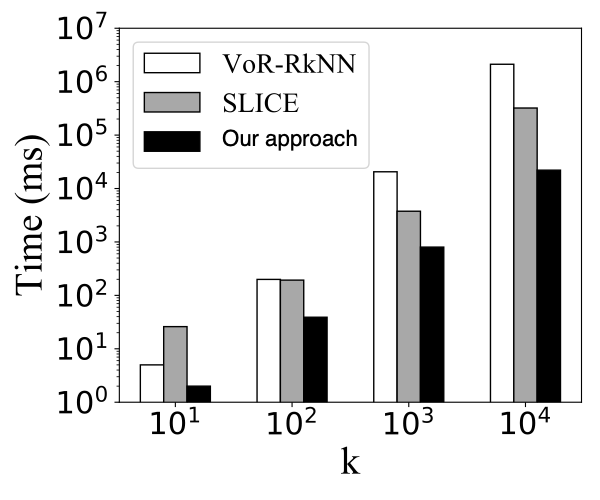}}
  \subfigure[Real data]{\includegraphics[width=0.235\textwidth]{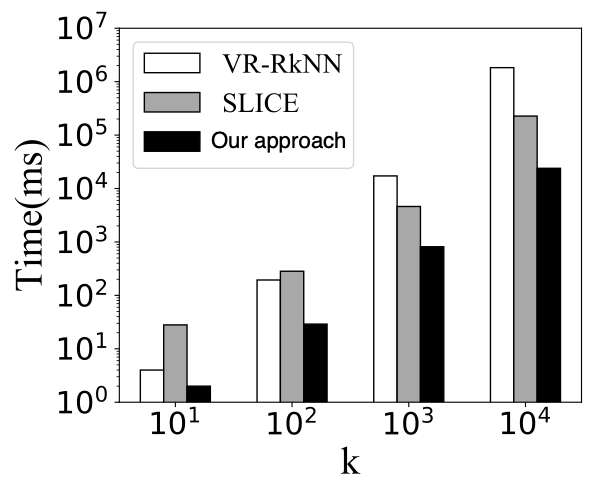}}
  \caption{Effect of $k$ on efficiency of R$k$NN queries}
  \label{fig:time_k}
\end{figure}
Figure~\ref{fig:time_k} shows the influence of $k$ on the efficiency of these three R$k$NN algorithms, where sub-figure (a) and (b) shows the time cost of R$k$NN queries from simulated data and real data, respectively.
As $k$ varies from 10 to 10,000, the time cost of these three algorithms increases.
With both synthetic data and real data, the query efficiency of our approach is significantly higher than that of the other two competitors. With the increase of $k$, this advantage becomes more and more obvious.
When $k$ is 10,000, the time cost of our approach is only about 1/10 of that of the state-of-the-art algorithm SLICE.
The detailed experimental results are presented in Table~\ref{table:time_k}.


\begin{figure}[htbp]
  \centering
  \subfigure[Simulated data]{\includegraphics[width=0.235\textwidth]{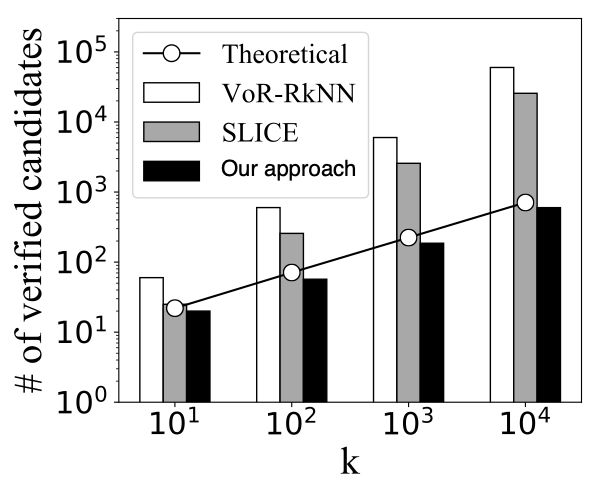}}
  \subfigure[Real data]{\includegraphics[width=0.235\textwidth]{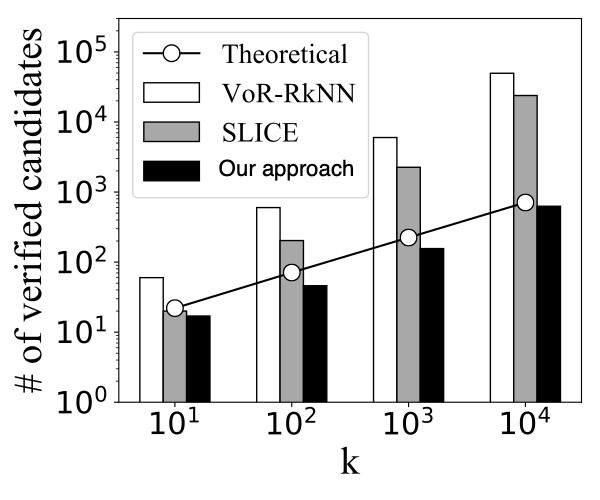}}
  \caption{Effect of $k$ on the number of candidates verified}
  \label{fig:verified_candidate_k}
\end{figure}
\begin{table}[htbp]
\centering
\caption{Number of candidates verified by R$k$NN algorithms with various values of $k$.}
\label{table:verified_candidate_k}
\begin{tabular}{@{}crrrrrr@{}}
\toprule
\multirow{2}{*}{\textbf{$k$}} & \multicolumn{3}{c}{\textbf{\small Simulated data}} & \multicolumn{3}{c}{\textbf{\small Real data}} \\ \cmidrule(lr){2-4} \cmidrule(l){5-7}
& {\fontsize{6.5}{0}\textbf {VR-R$k$NN}} & {\fontsize{6.5}{0}\textbf{SLICE}}  & {\fontsize{6.5}{0}\textbf{Our approach}}   & {\fontsize{6.5}{0}\textbf{VR-R$k$NN}}   & {\fontsize{6.5}{0}\textbf{SLICE}} & {\fontsize{6.5}{0}\textbf{Our approach}}\\ \midrule
$10^1$  & 60 & 25 & \textbf{20} & 60 & 20 & \textbf{17}\\
$10^2$  & 600 & 257 & \textbf{57} & 600 & 203 & \textbf{46}\\
$10^3$  & 6000 & 2572 & \textbf{186} & 6000 & 2257 & \textbf{156}\\
$10^4$  & 60000 & 25675 & \textbf{599} & 49601 & 23874 & \textbf{627}\\
\bottomrule
\end{tabular}
\end{table}
Figure~\ref{fig:verified_candidate_k} reflects the relationship between $k$ and the number of candidate points verified of the three algorithms in the experiments.
Sub-figure (a) and (b) show the experimental results on simulated data and real data, respectively.
These two sub-figures also show the theoretical number of candidate points verified with different values of $k$.
During the execution of our algorithm, only the points in the determine point set are verified by issuing $k$NN queries.
Therefore, the number of candidates verified is equal to the size of the determine point set.
As we discussed in section~\ref{discriminant_set_analysis},  the size of the determine point set is theoretically not larger than $7.1\sqrt{k}$.
In consequence, the theoretical number of verified candidates in Figure~\ref{fig:verified_candidate_k} is $7.1\sqrt{k}$.
It can be seen from the figure that the actual number of points verified is slightly less than the theoretical value, $7.1\sqrt{k}$. It indicates that the experimental results are consistent with our analysis.
It is also obvious from the figure that the number of verified candidate points of our approach is much smaller than that of the other two algorithms.
The detailed experimental results are presented in Table~\ref{table:verified_candidate_k}.

\begin{figure}[htbp]
  \centering
  \subfigure[Simulated data]{\includegraphics[width=0.235\textwidth]{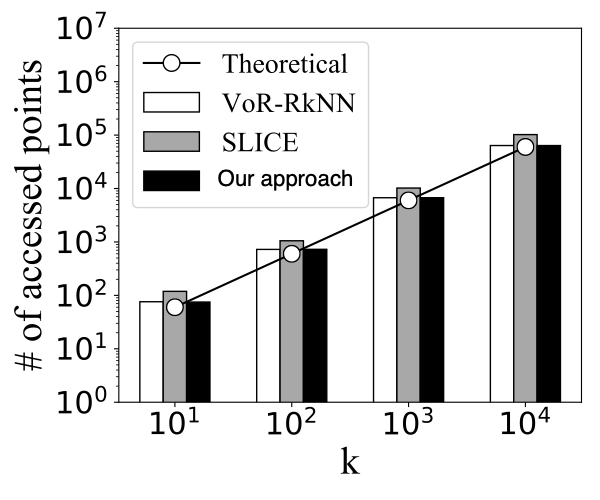}}
  \subfigure[Real data]{\includegraphics[width=0.235\textwidth]{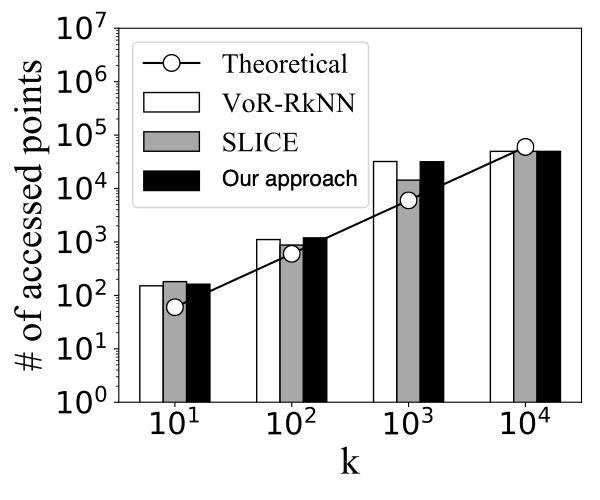}}
  \caption{Effect of $k$ on the number of points accessed}
  \label{fig:accessed_k}
\end{figure}
\begin{table}[htbp]
\centering
\caption{Number of accessed points of R$k$NN queries with various values of $k$.}
\label{table:accessed_k}
\begin{tabular}{@{}crrrrrr@{}}
\toprule
\multirow{2}{*}{\textbf{$k$}} & \multicolumn{3}{c}{\textbf{\small Simulated data}} & \multicolumn{3}{c}{\textbf{\small Real data}} \\ \cmidrule(lr){2-4} \cmidrule(l){5-7}
& {\fontsize{6.5}{0}\textbf {VR-R$k$NN}} & {\fontsize{6.5}{0}\textbf{SLICE}}  & {\fontsize{6.5}{0}\textbf{Our approach}}   & {\fontsize{6.5}{0}\textbf{VR-R$k$NN}}   & {\fontsize{6.5}{0}\textbf{SLICE}} & {\fontsize{6.5}{0}\textbf{Our approach}}\\ \midrule
$10^1$  & 76 & 119 & \textbf{75} & \textbf{153} & 181 & 162\\
$10^2$  & \textbf{725} & 1052 & 728 & 1108 & \textbf{876} & 1193\\
$10^3$  & \textbf{6721} & 10211 & 6731 & 32031 & \textbf{14359} & 31717\\
$10^4$  & 63782 & 102206 & \textbf{63721} & 49601 & 49601 & 49601\\
\bottomrule
\end{tabular}
\end{table}
Figure~\ref{fig:accessed_k} shows the number of accessed points of the three algorithms in the experiments and the theoretical number of accessed points of our approach with various values of $k$, which indirectly reflects their IO cost.
It can be seen from sub-figure (a), the number of accessed points of the three algorithms is almost equal in terms of magnitude, and so is the theoretical value of our approach. Specifically, the number of accessed points of our approach is slightly smaller than that of SLICE. As shown in sub-figure (b), our approach needs to access more points than SLICE. The reason is that the distribution of real data is very uneven, and our algorithm is more sensitive to the distribution of data than SLICE. Note that our approach and VR-R$k$NN use the same candidate set generation method, so they have almost the same number of accessed points.
The detailed experimental results are presented in Table~\ref{table:accessed_k}.

From the above three experiments, it can be seen that R$k$NN query efficiency is little affected by the data size, but greatly affected by the value of $k$.
Our approach is significantly more efficient than other algorithms because it requires less verification of candidate points.
For data sets with very uneven distribution of points, the candidate set of our approach is relatively large, which will affect the  IO cost to some extent.
However, the main time cost of the R$k$NN query is caused by a large number of verification operations rather than IO. Therefore, the distribution of points has little impact on the overall performance of our approach.

\section{Conclusions and future works}
In this paper, we propose an efficient approach to verify potential R$k$NN points without issuing any queries with non-constant computational complexity.
With the proposed verification approach, an efficient R$k$NN algorithm is implemented.
The comparative  experiments  are  conducted  between the proposed R$k$NN and other two R$k$NN algorithms of the state-of-the-art.
The experimental results show that our algorithm significantly outperforms its competitors in various aspects, except that our algorithm needs to access more points to generate the candidate set when the distribution of points is very uneven.
However, our algorithm does not require costly validation of each candidate point. 
Hence the distribution of data has very limited impact on its overall performance.
\bibliographystyle{unsrt}
\bibliography{paper}
\end{document}